\documentclass[letterpaper]{JHEP3}
\pdfoutput=1

\usepackage{amsmath}
\usepackage{amssymb}
\usepackage{amsfonts}
\usepackage{lscape, graphicx}
\usepackage{cite}       
\usepackage{bm}         
\usepackage{verbatim}

\usepackage[all]{xy}
\advance\parskip 1.0pt plus 1.0pt minus 2.0pt   
\def\href#1#2{#2}	

\def\R{{\mathbb R}}
\def\S{{\mathbb S}}

\def\tr{{\rm tr}}

\def\Z{{\mathbb Z}}

\def\Dslash{{\rlap{\raise 1pt \hbox{$\>/$}}D}}

\newcommand{\beq}{\begin{equation}}
\newcommand{\eeq}{\end{equation}}
\newcommand{\beqa}{\begin{eqnarray}}
\newcommand{\eeqa}{\end{eqnarray}}

\def\ltap{\ \raise.3ex\hbox{$<$\kern-.75em\lower1ex\hbox{$\sim$}}\ }
\def\gtap{\ \raise.3ex\hbox{$>$\kern-.75em\lower1ex\hbox{$\sim$}}\ }
\def\gl{\ \raise.5ex\hbox{$>$}\kern-.8em\lower.5ex\hbox{$<$}\ }
\def\roughly#1{\raise.3ex\hbox{$#1$\kern-.75em\lower1ex\hbox{$\sim$}}}

\title{ QCD  in magnetic field, Landau levels and  \\
double-life of unbroken center-symmetry}

\author
{
    {
    Mohamed M. Anber,$^1$\footnote{\email{manber@physics.utoronto.ca}}~
    Mithat {\"U}nsal,$^2$\footnote{\email{unsal.mithat@gmail.com}} 
           \\${}^{1}${Department of Physics, University of Toronto,
    Toronto, ON M5S 1A7, Canada}
    \\${}^{2}${Department of Physics and Astronomy, SFSU, San Francisco, CA 94132}
           
            }
    }%
    
    \abstract{

We study  the thermal confinement/deconfinement  and   non-thermal quantum  phase transitions or rapid cross-overs  in QCD and QCD-like  theories in external magnetic fields. 
     At large magnetic fields, while  the contribution of gauge fluctuations to Wilson-line   potential  remains  unaltered at one-loop order,  the contribution of fermions  effectively becomes  two lower dimensional  
          and  is enhanced by the density of states of the lowest Landau level (LLL). 
      In a  spatial compactification and for  heavy adjoint fermions, this enhancement leads to   a calculable  zero temperature quantum phase transition  on $ \mathbb R^3 \times \mathbb S^1$  driven by a competition between the center-destabilizing gauge contribution and 
      center-stabilizing LLL fermions.  We also show that at a (formal)   asymptotically large  magnetic field, the  adjoint fermions with arbitrarily large but fixed mass  stabilize  the  center symmetry. This is an exotic case of  simultaneous non-decoupling of large mass fermions (due to the enhancement by the LLL density of states) and decoupling from the   low energy effective field theory.  
      This observation has important implications  for both  Hosotani mechanism,   for which gauge symmetry  ``breaking" occurs,   and large-$N$  volume independence  (Eguchi-Kawai reduction),    for which gauge structure is  never ``broken".    Despite sounding almost self-contradictory, we carefully explain  the physical scales entering the  problem,  double-meaning of unbroken center symmetry and how a clash  is avoided.  We also identify,   for both thermal and spatial compactification, the jump in magnetic susceptibility   as an order parameter for the deconfinement transition.   
                 The  predictions of our analysis are testable by using current lattice techniques. 
    \smallskip
    
    \smallskip
    
       {\small{
     }

}}

\begin{document}

\maketitle

\section{Introduction}
Quarks carry both non-abelian color and abelian electric charges. In relativistic heavy ion collisions (RHIC),  large external $U(1)_{\rm em}$ magnetic field (of order $\sqrt {|eB|}  \sim 10^2$ MeV)  is generated. This is parametrically of order QCD-strong scale. 
Therefore, it is of experimental interest  to study both equilibrium  thermodynamics and non-equilibrium properties of QCD in external $B$-fields.

  A magnetic field introduces a Landau level structure to the  fermion
spectrum.   Few rather interesting phenomena stem from this:  chiral magnetic effect  which is an interplay of the LLL structure and topological 
aspects of QCD \cite{Kharzeev:2004ey,Kharzeev:2007tn,Kharzeev:2007jp,Fukushima:2008xe,Basar:2012gm} and magnetic catalysis  which helps spontaneous breaking of non-abelian  chiral symmetry even at very weak coupling \cite{Gusynin:1995nb,Gusynin:1994re,Gusynin:1999pq}, 
inverse magnetic catalysis  and non-monotoniticity observed in lattice simulations \cite{ Bali:2011qj, Bali:2012zg, Bali:2013esa, Bruckmann:2013oba}.  Also see  \cite{D'Elia:2010nq, D'Elia:2011zu} for simulations of QCD in external $B$-field.  

In this work,  our goal is to study the 
 the role of the  $B$-fields in center-symmetry realization, and   the equilibrium thermodynamics and some aspects of phase structure for  QCD-like theories, with fermions in one and two-index  representations ${\cal R}$. An interesting question  is whether the back-reaction of the fermions in varying-$B$ field   can  alter the  phase of the theory, say, from a center-broken phase to a center symmetric phase or vice versa.  We find an example of such phenomena for adjoint   representation  fermions.  Another interesting question is 
 the interplay of external-  $U(1)_{\rm em}$ $B$-fields,   monopole-instantons  (with fractional topological charge) which carry chromomagnetic $B$-field, and chiral symmetry realization which we study in a follow-up.

We  study center-symmetry realization in  both thermal  and spatial compactification in the presence of  external magnetic fields. 
 In path integral formalism, integrating out fermions with anti-periodic  (periodic) spin connection correspond to the thermal (twisted) partition function. In operator formalism, this amounts to regular (graded)  trace  over the Hilbert space, namely 
 \begin{align} 
   Z_\eta&= Z_{\cal B} - \eta Z_{\cal F}   =     \tr (e^{- L H} (-\eta)^F )   \cr
 & =\int_{A_{\mu} (L ) = A_{\mu} (0 ) }   DA_{\mu} \;  e^{-S[A]} 
  \int_{\psi(L ) =  \eta  \psi(0) } D \psi D \bar \psi  \; e^{  \int_{\mathbb R^3\times \mathbb S^1_\eta}\bar \psi \left(-i \displaystyle{\not} D+m  \right)  \psi}   \cr
&=  \int_{A_{\mu} (L ) = A_{\mu} (0 ) } DA_{\mu} \;  e^{-S[A]}   
 \;    {\rm det}_{\eta}  \left(-i \displaystyle{\not} D+m  \right)\,,  \cr  
  \eta&=  \left\{ \begin{array}{lll}  - & 
\qquad  {\rm   thermal  \; circle,}\,\, \mathbb S^1_- & \qquad  L= {\rm  \;  \beta }= 1/T\,,  \\ 
  + &  \qquad {\rm spatial \; (non-thermal) \; circle,}\,\, \mathbb S^1_+ &  \qquad  L =  L \,,
  \end{array} \right.
  \label{part-func}
 \end{align} 
where  $(-1)^F$ is fermion number modulo two, acting as $\pm$ on bosonic (fermionic) Hilbert spaces,  
 and ${\rm det}_{\mp} $ corresponds to the determinant in the space of anti-periodic/periodic functions.\footnote{ It is important to note that the periodic boundary condition for fermions is {\it  not} unphysical, it has a well-defined meaning in operator formalism.   In either case, fermions  are spin-half particles and they  obey the Pauli exclusion principle,  and anti-commutation relations.  However,  the spatial compactification  does not have a thermal interpretation, and the Fermi-Dirac distribution  (relevant to thermal QFT) only arise in the thermal compactification.}   
 By studying the properties of the Dirac operator  $  (-i \displaystyle{\not} D+m)$ 
  in the presence of external magnetic field and a background Wilson line, we find the fermion induced  one-loop potential.\footnote{
 We take  the $U(1)_{\rm em}$ magnetic field as an external field, with no dynamics associated with it. Otherwise, at small-$L$ and vanishing fermion mass, the abelian part would  be strongly coupled.}  
 Calculationally, this is  a standard generalization of the   Euler-Heisenberg effective Lagrangian (see e.g. \cite{Dunne:2012vv, Dunne:2004nc} and references therein) and   Gross-Pisarski-Yaffe one-loop potential for Wilson line \cite{Gross:1980br,Shore:1981mj, Cangemi:1996tp, Dittrich:1979ux,Braden:1981we, Meisinger:2001fi, Meisinger:2002ji, Unsal:2006pj, Mizher:2010zb}.  We express the fermion induced Wilson line potential  as a sum over Landau level contributions.

 Consider QCD   with gauge group $G$ with fermions in representation ${\cal R}$, where fermions also carry charges under 
 $U(1)_{\rm em}$.    For ${\cal R}$, we primarily consider  $n_f$  fundamental (F), anti-symmetric (AS),  symmetric (S) Dirac fermions  and $n_f$ adjoint (adj) Weyl fermions.   For adjoint matter, we only consider  $n_f=$  even so that  we can build $n_f/2$ Dirac fermions to which we can assign an electric charge without causing any gauge anomaly. The motivation to study two-index representation is that 
 QCD(AS) is  a natural generalization of ordinary QCD to large-$N$,  and it is related via orientifold equivalence 
  to the adjoint representation \cite{Armoni:2003gp, Armoni:2003fb,  Unsal:2006pj}.

\vspace{0.3 cm}
{ \bf Thermal compactification and phase structure:} 
At strong magnetic fields, the fermion induced Wilson line potential  is dominated by the lowest Landau level (LLL), and undergoes dimensional reduction by 
two dimensions, similar to the chiral condensate  \cite{Gusynin:1995nb,Gusynin:1994re,Gusynin:1999pq}: \begin{eqnarray}
{\cal V}_{-} [\Omega]&=&  
{\cal V}_{\mbox{\scriptsize gauge}}^{\mathbb R^3 \times S^1_\beta} [\Omega] +   {\cal V}_{ - , {\cal R}}^{\mathbb R^3 \times S^1_\beta} [\Omega]   
\qquad \xrightarrow{{\rm large-B}} \qquad 
 {\cal V}_{\mbox{\scriptsize gauge}}^{\mathbb R^3 \times S^1_\beta} [\Omega] + \left(\frac{eB}{2\pi} \right)  {\cal V}_{ - , {\cal R}}^{\mathbb R^1 \times S^1_\beta} [\Omega]\,,  \qquad 
\label{thermal-EHGPY} 
\end{eqnarray}
where  $ \left(\frac{eB}{2\pi} \right) $ is the density of states of the LLL.     
${\cal V}_{\mbox{\scriptsize gauge}}^{\mathbb R^3 \times S^1_\beta} [\Omega] $
is the standard contribution of gauge fluctuation  to the Wilson line potential \cite{Gross:1980br}.   For fermions,  life becomes essentially two dimensional. If ${\vec B}  = B \hat z $, then, one effectively  deletes the $xy$-plane and the fermions are localized to the two-dimensional $zt$ plane. Furthermore,  their effect  is parametrically  enhanced by
 $ \left(\frac{eB}{T^2} \right) =  \left(\frac{\beta}{\ell_m} \right)^2  \gg 1$ where  $\ell_m \sim 1/ \sqrt {eB}$ is the magnetic length scale.    
  Extremizing the potential yields free energy density, given by 
\begin{eqnarray}
{\cal F} =   -  {\rm dim}({\rm adj})  \times \underbrace{ \left( \frac{\pi^2}{45}  T^4 \right) }_{\rm Stefan-Boltzmann \;  4d}  -   n_f   {\rm dim}({\cal R} )   \times  \underbrace{ \left(\frac{|eB|}{2\pi}  \right)  }_{\rm LLL \; density \;of\; states} \times  \underbrace { \left( \frac{\pi}{12}  T^2 \right) }_{\rm Stefan-Boltzmann \;  2d}  
\end{eqnarray}
in accordance  with the LLL interpretation and dimensional reduction. 
 
 \vspace{0.3 cm} 
 { \bf Spatial  compactification and a quantum phase transition:} 
 An interesting   gauge phenomenon occurs for ${\cal R} = \rm adj$ where fermions (with mass $m$)  are endowed with periodic boundary conditions, $\eta= +1$ in   (\ref{part-func}).
   \begin{FIGURE}[ht]
    {
    \parbox[c]{\textwidth}
        {
        \begin{center}
        \includegraphics[angle=0, scale=.40]{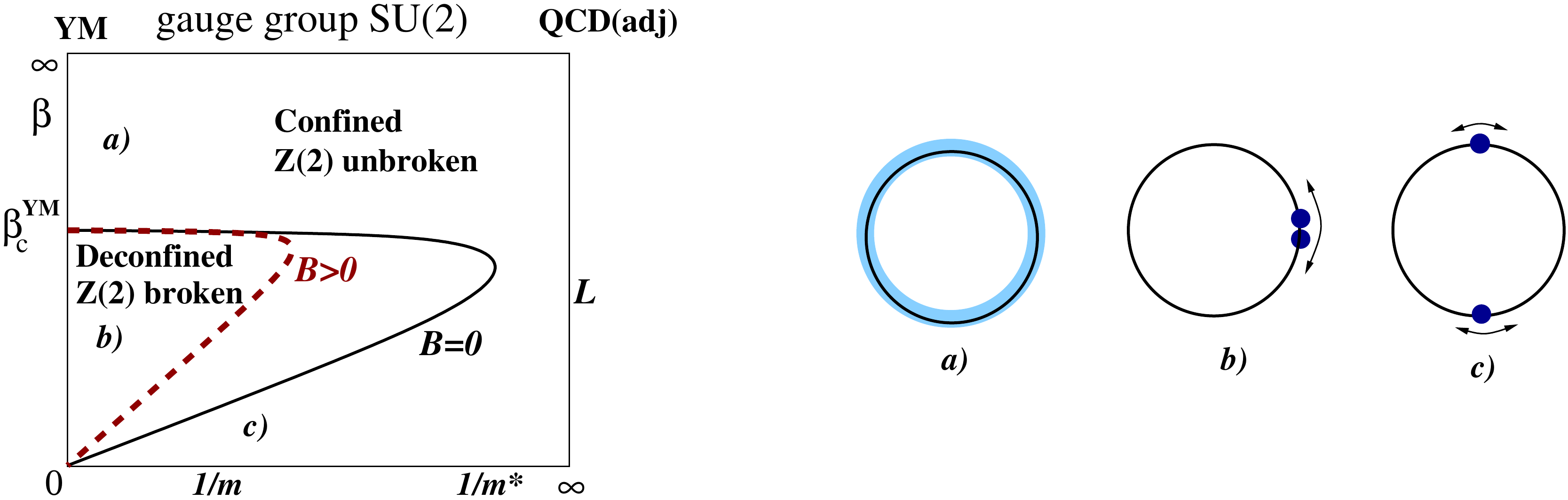}
	\hfil
        \caption 
      {Left: The phase diagram for $SU(2)$ gauge theory with $n_f=2$  adjoint Weyl fermions, on $\mathbb R^3 \times \mathbb S^1_+$, in the $L$-$1/m$ plane for $B=0$ and $|B|>0$.  Center-broken regime shrinks with increasing magnetic field. 
      Right: a) ``Cartoon"  of strong coupling non-trivial holonomy $\langle {\rm tr} \Omega \rangle=0 $, eigenvalues are randomized over the unit circle.  b) Weak coupling  trivial holonomy $\langle {\rm tr}\Omega \rangle=1$. c) Weak  coupling  non-trivial holonomy $\langle {\rm tr} \Omega \rangle=0 $, eigenvalues are at anti-podal points, and the fluctuations in their position is small.       
       a) and c) domains are both center-symmetric and are  continuously connected. 
			}
      \label{Phse diagram}
       \end{center}
        }
    }
\end{FIGURE}
When the fermions are massless or  sufficiently  light, they induce a center-stabilizing potential,  leading to gauge ``symmetry breaking" or 
  adjoint Higgsing or  abelianization  at one loop-order, and in fact, to all orders in perturbation theory. 
  
  This result has two mutually  independent  and exclusive  histories. One is in the context of   gauge-Higgs unification     \cite{Hosotani:1988bm, Higuchi:1988qa, Davies:1987ei}  for which gauge symmetry breaking (Hosotani mechanism)  {\it occurs} and the other is in the discussion of    large-$N$ volume independence    \cite{Kovtun:2007py, Unsal:2007jx}   
  (working realization of Eguchi-Kawai reduction\cite{Eguchi:1982nm})   where gauge symmetry breaking {\it never occurs}, and   
the    semi-classical calculable regime   where abelianization again {\it occurs} \cite{ Unsal:2007jx}.   The discussion of  scales, the role of the parameter $ \frac{LN\Lambda}{2\pi} $ which determines whether a center-symmetric regime exhibits   adjoint Higgsing  or not    first appeared more recently in    \cite{Unsal:2008ch, Unsal:2007jx},  in distinguishing    large-$L$  or large-$N$ (gauge structure unbroken) and small-$LN$   adjoint Higgsing semi-classical calculable  regimes, and did not appear in earlier work. 
In particular, 
   abelianization
   and semi-classical calculability takes place in the $\frac{LN\Lambda}{2\pi}  \lesssim 1$ domain \cite{Unsal:2007jx}, and large-$N$ volume independence in the  $\frac{LN\Lambda}{2\pi}  \gg 1$ domain \cite{Kovtun:2007py, Unsal:2010qh}.  
  (Also see   more recent works \cite{Armoni:2011dw, Perez:2013dra}   emphasizing the role of $LN \Lambda$ parameter, and recent reviews  of large-$N$ limits \cite{Lucini:2012gg,Lucini:2013qja}.)
  As explained in detail in Section \ref{sec:abel}, the discussion of scales clarifies  how a contradiction is avoided between these two different regimes.    
The understanding of the role of parameter  $\frac{LN\Lambda}{2\pi}$  is extremely important in finding lattice realization of these two regimes.

  When the fermions are heavy, this theory has an exotic phase structure, shown in Fig.~\ref{Phse diagram}, 
 center-symmetric at sufficiently small and sufficiently large $S^1$, and center-broken in between  \cite{Unsal:2010qh, Cossu:2009sq}.\footnote{The lattice  simulations  in   \cite{Cossu:2009sq}  exhibits  the existence of small and large-$L$ confined phases, but do not currently show their continuity 
 on the small mass regime $m<m^*$. 
   However, there is strong theoretical reasons to believe that the theory will not have center-broken intermediate regime for $m<m^*$.  The reason for the  non-observation in   \cite{Cossu:2009sq}    may be that the simulations are not run at sufficiently light fermion masses.} 
 This system is interesting because it does not have a  strict thermal interpretation, but it admits a non-thermal quantum phase transition.  Its phase diagram in the $L$-$1/m$  plane in the absence of magnetic field  is studied in \cite{Unsal:2010qh}.  We study the same phase diagram in the presence of large-$B$ fields. As shown in  Fig.~\ref{Phse diagram},   the center-broken regime shrinks with the application of the $B$-field. This happens when the magnetic field is sufficiently  large such that it can  compensate suppression due to the  mass term  for fermion. 
This is a rather exotic phase transition driven by the competition between center-destabilizing gauge fluctuations and the increase of the LLL density of states of the adjoint fermions endowed with periodic boundary conditions.  This transition can be checked by using standard lattice simulations, by adding magnetic field to the set-up of  \cite{Cossu:2009sq}.

\section{Turning on magnetic field in QCD on $\mathbb R^3 \times \mathbb S^1$ }

We consider $SU(N)$ gauge theory coupled to massive fermions on $\mathbb R^3 \times \mathbb S^1$ which obey either periodic or anti-periodic boundary conditions along $\mathbb S^1$. We couple the fermions to a  background $U(1)_{\rm em}$ gauge field that is taken to be constant and  perpendicular to the $\mathbb S^1$ circle.  In the following, it will prove easier to work with Dirac fermions.
Hence, the system Lagrangian reads
\begin{eqnarray}
{\cal L}=-\frac{1}{4g^2}F_{\mu\nu}^aF^{a\,\,\mu\nu}+\bar\psi\left(\displaystyle{\not}\partial-i\displaystyle{\not}A^a T^a+ie\displaystyle{\not}A_{\mbox{\scriptsize em}}+m\right)\psi\,,
\end{eqnarray}
where $e$ is the electromagnetic coupling constant and $T^a$ are the Lie generators in the appropriate representation ${\cal R}$. Next, we analytic continue to the Euclidean space and integrate out the fermions to obtain the one-loop effective action $\Gamma^{\mbox{\scriptsize Dirac}}=\mbox{Tr}\log \left(-i \displaystyle{\not} D+m  \right)$, where  $\displaystyle{\not} D=\displaystyle{\not}\partial-i\displaystyle{\not}A^a T^a+ie\displaystyle{\not}A_{\mbox{\scriptsize em}}$, and $\mbox{Tr}$ denotes the trace over spacetime, Dirac and color indices. Using the fact that the sign of the fermion mass is irrelevant, we get
\begin{eqnarray}
\Gamma^{\mbox{\scriptsize Dirac}}=\mbox{Tr}\log \left(-i \displaystyle{\not} D+m  \right)= \mbox{Tr}\log \left(i \displaystyle{\not} D+m  \right)=\frac{1}{2}\mbox{Tr}\log\left(-\displaystyle{\not} D^2+m^2\right)\,,
\end{eqnarray}
where $\displaystyle{\not} D^2=D^2-\sigma_{\mu\nu}\left(F^{a\,\,\mu\nu}T^a+eF_{\mbox{\scriptsize em}}^{\mu\nu}\right)/2$, and $\sigma_{\mu\nu}=\frac{i}{2}\left[\gamma_\mu,\gamma_\nu\right]$. The effective action  $\Gamma^{\mbox{\scriptsize Dirac}}$ is a divergent quantity. Therefore, we regularize it by subtracting out the free field contribution:
\begin{eqnarray}
\Gamma^{\mbox{\scriptsize Dirac}}_{\mbox{\scriptsize reg}}=\frac{1}{2}\mbox{Tr}\log\frac{\left(-\displaystyle{\not} D^2+m^2\right)}{\left(-\Box+m^2\right)}\,,
\end{eqnarray}
such that we have $\Gamma^{\mbox{\scriptsize Dirac}}_{\mbox{\scriptsize reg}}=0$ as we turn off both the color and electromagnetic fields.

In general, the calculation of $\Gamma^{\mbox{\scriptsize Dirac}}_{\mbox{\scriptsize reg}}$ is a formidable task. However, it turns out that this problem can have an exact solution in a few special cases. We specify our problem by turning on a constant holonomy (or Wilson line) $A_0^a$ along the $\mathbb S^1$ direction and ignoring the gauge fluctuations in all other directions. In consequence, the chromo-field strength vanishes $F^{a\,\,\mu\nu}=0$, i.e.  the non-abelian gauge connection is flat. Then, using the integral representation of the $\log$ function, we obtain 
\begin{eqnarray}
\nonumber
\Gamma^{\mbox{\scriptsize Dirac}}_{\mbox{\scriptsize reg}}&=&\frac{1}{2}\mbox{Tr}\log\frac{-D^2- e\sigma\cdot F_{\mbox{\scriptsize em}}/2+m^2}{-\Box+m^2}\\
&=&-\frac{1}{2}\mbox{Tr}\int_0^\infty \frac{d\tau}{\tau}\left(e^{-\tau\left(-D^2- e\sigma\cdot F_{\mbox{\scriptsize em}}/2+m^2\right)}-e^{-\tau\left(-\Box+m^2\right)}\right)\,,
\end{eqnarray}
where $D^2=(\partial_0+A_0^aT^a)^2+\left(\partial_i+ ieA_{\mbox{\scriptsize em}\,\,i}\right)^2$. The trace over the free field part is trivial and can be performed directly by  going to the momentum space.  Since the electromagnetic field is assumed to be perpendicular to $\mathbb S^1$, one can break the trace into two independent parts: one along the compact dimension and the other along the infinite dimensions as follows: 
\begin{eqnarray}
\nonumber
\Gamma^{\mbox{\scriptsize Dirac}}_{\mbox{\scriptsize reg}}&=&-\frac{1}{2}\sum_{n\in \mathbb Z}\int_0^\infty \frac{d\tau}{\tau}e^{-m^2\tau}\left\{\mbox{tr}_{\cal R}\left[e^{-\tau\left(\omega_n+A_0^aT^a\right)^2}\right]\times\mbox{tr}\left[e^{-\tau\left[\left(\partial_i+i eA_{\mbox{\scriptsize em}\,\,i}\right)^2+ e \sigma\cdot F_{\mbox{\scriptsize em}}/2\right]} \right]\right.\\
&&\left.\quad\quad\quad\quad\quad\quad\quad\quad\quad-4\int \frac{d^3 k}{\left(2\pi\right)^3}e^{-\tau\left[(\omega_n^2+ k^2)+m^2\right]}\right\}\,,
\end{eqnarray}
where $\omega_n$  are the Matsubara frequencies which are given by $2\pi n/L$ and $(2n+1)\pi/L$ for spatial and thermal compactifications, respectively. During this process of integrating out the Kaluza-Klein modes, we have assumed that these modes are weakly coupled and can be integrated out perturbatively.  This assumption is justified in the small-$L$ regime by the  asymptotic freedom of QCD, for sufficiently  small number of fermions. (Recall that the $U(1)_{\rm em}$ magnetic field is treated as a background, with no dynamics associated with it. Otherwise, at small-$L$, the abelian part would  be strongly coupled.)  
The first trace $\mbox{tr}_{\cal{R}}$ is over the Lie algebra representation ${\cal R}$, while the second trace is over space and Dirac indices. The trace $\mbox{tr}\left[e^{-\tau\left[\left(\partial_i+i eA_{\mbox{\scriptsize em}\,\,i}\right)^2+ e \sigma\cdot F_{\mbox{\scriptsize em}}/2\right]} \right]$ is a standard Euler-Heisenberg calculation which encodes information about a constant electromagnetic field in $3$ dimensions. Turning on only the magnetic field, setting the electric field to zero, we have
\footnote{In the case of spatial compactification, the magnetic field has only a single component in  $\mathbb R^{3}$  (recall that in this case one of the dimensions in $\mathbb R^3$ is the time dimension; in $2+1$ dimensions the magnetic field has only one component). On the other hand, in the case of thermal compactification the magnetic field can have three components. However, we can always choose the magnetic field to be aligned in the $\hat z$-direction. Thus, (\ref{the infinite dimension contribution}) is valid for both spatial and thermal compactifications. }
\begin{eqnarray}
\mbox{tr}\left[e^{-\tau\left[\left(\partial_i+i eA_{\mbox{\scriptsize em}\,i}\right)^2+ e \sigma\cdot F_{\mbox{\scriptsize em}}/2\right]} \right]=4\frac{V_{\mathbb R^3}}{\left(4\pi\tau\right)^{3/2}}\frac{e\tau B}{\tanh\left(e\tau B\right)}\,,
\label{the infinite dimension contribution}
\end{eqnarray}
where $V_{\mathbb R^3}$ is the three dimensional volume. Putting things together we find
\begin{eqnarray}
\nonumber
\Gamma^{\mbox{\scriptsize Dirac}}_{\mbox{\scriptsize reg}}&=&-2\sum_{n\in Z}\int_0^\infty \frac{d\tau}{\tau}\frac{V_{\mathbb R^3}}{\left(4\pi\tau\right)^{3/2}}e^{-m^2\tau}\left[\mbox{tr}_{\cal R}\left(e^{-\tau\left(\omega_n+A_0^aT^a\right)^2}\right)\times\frac{e\tau B}{\tanh\left(e\tau B\right)}\right.\\
&&\left.\quad\quad\quad\quad\quad\quad\quad\quad\quad\quad\quad\quad\quad-e^{-\tau\omega_n^2}\right]\,.
\end{eqnarray} 

At this stage, we define the effective potential ${\cal V}$ as ${\cal V} \equiv -\Gamma/(LV_{\mathbb R^3})$. Using the Poisson resummation formula 
\begin{eqnarray}
\sum_{n \in Z}e^{-\tau \left(\omega_n+q\right)^2}=\frac{L}{\sqrt{4\pi \tau}}\sum_{n \in Z}e^{-\frac{L^2n^2}{4\tau}+inLq}\,,
\end{eqnarray}
and the change of variables $\tau=L^2 y$, we obtain the effective potential per Dirac fermion
\begin{eqnarray}
{\cal V}^{\mathbb R^3\times \mathbb S^1_\pm}
=\frac{2}{\left(4\pi\right)^2L^4}\sum_{n=1}^{\infty}\int_{0}^{\infty}\frac{dy}{y^3}e^{-\frac{n^2}{4y}-m^2L^2y}\left\{\frac{e BL^2y}{\tanh \left(e BL^2 y\right)}a_n\left(\mbox{tr}_{\cal R}\Omega^n+\mbox{c.c.}\right)-2\right\}\,,
\label{the main result of the work}
\end{eqnarray}
where
\begin{eqnarray}
\Omega=e^{iLA_0^aT^a}
\end{eqnarray}
is the Wilson line wrapping the $\mathbb S^1$ circle, and the pre-factor $a_n$ is
\begin{eqnarray}
a_n=\left\{
\begin{array}{cc}
(-1)^n & \mbox{for thermal compactification}\,\, \mathbb S^1_-\,, \\
1 & \mbox{for spatial compactification}\,\, \mathbb S^1_+\,.
\end{array}
 \right.
\end{eqnarray}
depending on the spin-connection of fermions over the $\mathbb S^1$ circle. 
 Notice that in obtaining (\ref{the main result of the work}) we omitted  the zero mode, $n=0$, which gives a divergent but otherwise   holonomy independent contribution.
\footnote{ The $n=0$ term corresponds to the fermions vacuum correction in the presence in the magnetic field, and leads to charge renormalization which we ignore here.}
 We also note that the last term in (\ref{the main result of the work}) is independent of $B$ and $A_0^a$ and hence can be neglected in our subsequent analysis. Finally, upon using the change of variables $u=m^2L^2y$ in (\ref{the main result of the work}),  we find
\begin{eqnarray}
{\cal V}^{\mathbb R^3\times \mathbb S^1_\pm}=\frac{2}{\pi^2L^4}\sum_{n=1}{\cal M}^2_n(m,B)a_n\frac{\left(\mbox{tr}_{{\cal R}}\Omega^n+\mbox{c.c.}\right)}{n^4}\,,
\label{the final form of the effective potential}
\end{eqnarray}
where the effective mass square term ${\cal M}^2_n(m,B)$ is given by
\begin{eqnarray}
{\cal M}^2_n(m,B)=\frac{z_n^4}{16}\int_0^\infty \frac{du}{u^3}e^{-\frac{z_n^2}{4u}-u}\frac{xu}{\tanh (xu)}\,,
\label{exact result}
\end{eqnarray}
and $z_n=nmL$, and $x=eB/m^2$. Equation (\ref{the final form of the effective potential}) is our main result. The form of $\mbox{tr}_{\cal{R}}\Omega^n+\mbox{c.c.}$ for the fundamental (F), adjoint (adj), symmetric (S) and anti-symmetric (AS) representations is given by  \footnote{
For $SU(N)$  pure YM and QCD(adj),  the center symmetry is $\Z_N$. For odd $N$, and ${\cal R}$= F, S/AS, the center symmetry is trivial, $\Z_1$. For even $N$, ${\cal R}$= F, S/AS, the center symmetry is $\Z_1$, and $\Z_2$. These global symmetries are also manifest in the one-loop potential.}
\begin{eqnarray}
\mbox{tr}_{{\cal R}}\Omega^n+\mbox{c.c.}=
\left\{
\begin{array}{ll}
\mbox{tr}\Omega^n+\mbox{tr}\Omega^{*n}\,, & \;\; \mbox{1-F Dirac}\,,\\
\left|\mbox{tr}\Omega^n\right|^2\,,&  \;\; \mbox{1-adj Weyl}\,,\\
\frac{1}{2}  \left[ (\mbox{tr} \Omega^n)^2  \mp \mbox{tr} \Omega^{2n}    \right]+ \mbox{c.c.}
\,,& \;\;  \mbox{1-AS/S Dirac}\,.
\end{array}\right.
\label{summary of all representations}
\end{eqnarray}
Notice that we give the result per Weyl fermion for the case of adjoint representation, keeping in mind that in this case we need an even number of Weyl fermions to avoid gauge anomaly.  In the $B=0$ and $m=0$,  and $B=0$ and $m\neq0$,  we obtain known  results in thermal   \cite{Gross:1980br, Meisinger:2001fi, Meisinger:2002ji,  
Unsal:2006pj}   and spatial compactification \cite{Hosotani:1983xw, Unsal:2006pj,Argyres:2012ka}, also see \cite{ Kashiwa:2013rmg, Kouno:2013mma}.  
In the large magnetic field limit, this expression reduces to 
\begin{eqnarray}
{\cal V}_{\pm} [\Omega]& \xrightarrow{{\rm large-B}}&
{\cal V}_{\mbox{\scriptsize gauge}}^{\mathbb R^3 \times S^1_L} [\Omega] + \left(\frac{|eB|}{2\pi} \right)  {\cal V}_{ \pm ,}^{\mathbb R^1 \times S^1_L} [\Omega]  \cr
&=& 
-\frac{2}{\pi^2L^4}\sum_{n=1}\frac{\left|\mbox{tr}\Omega ^n \right|^2}{n^4} + 
 \left(\frac{|eB|}{2\pi} \right)    \frac{n_f }{\pi L^2} \sum_{n=1}\frac{ (nLm) K_1(nLm) } {n^2} (\pm)^n \left(  \mbox{tr}_{{\cal R}}\Omega^n+\mbox{c.c.} \right)\,. \qquad 
\label{result for strong field}
\qquad 
\end{eqnarray}
Hence, in the presence of a strong magnetic  field,  the fermion contribution  behaves as if fermions  live on a space-time dimensionality $d-2=2$, i.e. on $\mathbb R^1 \times 
\S^1$ instead of  $\mathbb R^3 \times 
\S^1$.

\section{Landau levels and the role of the lowest Landau level}
In this section, we express  the one-loop potential    (\ref{the main result of the work})
as a sum over all  Landau levels. In particular, 
 we show that   the strong field limit (\ref{result for strong field}) is solely due to the contribution of the lowest Landau level (LLL).

The spectrum of the Dirac operator   on $\mathbb R^3\times \mathbb S^1$ in the presence of a non-trivial holonomy along the $\mathbb S^1$ direction and magnetic field $B$ perpendicular to $\mathbb S^1$ is given by
\begin{eqnarray}
\lambda_{\sigma,p,n,k_z,A_0}=m^2+k_z^2+\left(\omega_n +A_0^aT^a\right)^2+|eB|(2p+1 + \sigma)\,,
\end{eqnarray}
where $k_z$ is the momentum along the $z$-direction (perpendicular to both $\mathbb S^1$ and the $x-t$ plane), $\omega_n$ is the Kaluza-Klein frequency along the compact direction,   $p=0, 1, 2, \ldots$ is the Landau level, $\frac{\sigma}{2} = \frac{\pm 1}{2}$ is  the spin.
Every Landau level has $|eB|/(2\pi)$ degeneracy factor for each spin alignment.   Note that  the LLL  is given by $p=0, \sigma= -$, 
while the higher Landau levels also have additional pairing degeneracy between   $(p+1, \sigma= -)$ and $(p, \sigma= +)$.

 The zeta function associated with the Dirac operator is given by\footnote{
Recall that the determinant of an operator ${\cal O}$  with eigen-spectrum $\{\lambda\}$, i.e.,  ${\cal O}\psi_\lambda=\lambda\psi_\lambda$, 
is given by
$\mbox{Det} {\cal O}=\prod_\lambda \lambda=e^{\sum_\lambda \log\lambda}=e^{{\rm tr} \log {\cal O}}$, where $\{\lambda\}$ are the eigenvalues of the operator $O$.  
Using the definition of the zeta function, $\zeta(s)=\sum_\lambda \lambda^{-s}$,
we find
$\mbox{Det} {\cal O}=\exp[-\zeta'(s=0)]\,.$
}
\begin{eqnarray}
\zeta_{\mbox{\scriptsize Dirac}}(s)=V_{\mathbb R^3}\frac{|eB|}{2\pi}\sum_{p=0}^{\infty} \sum_{n\in \mathbb Z}\sum_{\sigma=\pm}\int \frac{dk_z}{2\pi}\mbox{tr}_{{\cal R}}   \left[  (\lambda_{\sigma,p,n,k_z,A_0})^{-s} \right]  \;  .
\end{eqnarray}
The fermion contribution to the  one-loop potential   ${\cal V}$  for the Wilson line holonomy   on  $\mathbb R^3\times \mathbb S^1$ 
can be extracted from this expression and is given by  
\begin{align}
{\cal V}[\Omega] = -\log Z / 
(LV_{\mathbb  \R^3})= \zeta'(0)/(LV_{\mathbb  \R^3})\,,
\end{align} 
where  the logarithm of partition function is 
\begin{eqnarray}
\log Z= 
V_{\mathbb R^3}\frac{|eB|}{2\pi}\sum_{p=0}^{\infty} \sum_{n\in \mathbb Z}\sum_{\sigma=\pm}\int \frac{dk_z}{2\pi}\mbox{tr}_{{\cal R}}   \left[   \log \lambda_{\sigma,p,n,k_z,A_0}  \right]  \;  .
\label{part-QFT}
\end{eqnarray}

Before proceeding with this expression, it is also useful to make connection with the usual methods of statistical mechanics. 
 The partition function of a  free fermion gas in a magnetic field  is  $Z=  {\rm tr} ( e^{-\beta H}) =   \prod_Q  Z_Q= \prod_Q  (1+ e^{- \beta E_Q})^{-1} $,  where $Q= \{   p, k_z, \sigma\}$ is a collective index for the quantum numbers of the states (defined above), and 
$\log Z =  - \sum_Q \log  (1+ e^{- \beta E_Q})  $. The energy eigenstates for a relativistic particle in a constant magnetic field is 
given by  
\begin{align}
E_{k_z, p, \sigma}    = \sqrt {m^2 + k_z^2 + |eB| (2 p+1 + \sigma) }\,.
\end{align}
Consequently, the partition function can be written as  
\begin{align}
-\log Z =  2 {\rm dim} ({\cal R})  V_{\mathbb R^3} \frac{|eB|}{2\pi} \sum_{p=0}^{\infty}  \sum_{\sigma=\pm} \int \frac{dk_z}{2\pi}   \log  (1+ e^{- \beta E_{k_z, p, \sigma}})\,.    
\label{part} 
 \end{align}
 If a Wilson line  $\Omega =e^{i \beta  A_0^aT^a}$  is turned on, this expression is modified into 
\begin{align}
-\log Z =    V_{\mathbb R^3} \frac{|eB|}{2\pi} \sum_{p=0}^{\infty}  \sum_{\sigma=\pm} \int \frac{dk_z}{2\pi}  \left[  {\rm tr}_{\cal R}  \log  (1+ e^{- \beta E_{k_z, p, \sigma} } \Omega  )  + {\rm c.c.} \right]\,, 
\label{part-W} 
 \end{align}
where the first term is due to quarks and the second term is due to anti-quarks. For trivial Wilson line background, i.e.  $A_0^aT^a=0$, 
(\ref{part-W}) reduces to (\ref{part}).   

 In the field theory expression (\ref{part-QFT}), 
performing the sum  over the Kaluza-Klein modes  gives the  statistical mechanics expression (\ref{part-W}),  and this reduces to   (\ref{part}) for trivial holonomy background.

Using the degeneracy $E_{k_z, p+1, -} = E_{k_z, p, +}  $  for $p \geq 1$, we can perform the summation over spin $\sigma$, and rewrite
(\ref{part-W})  as a  sum over the Landau levels, where the LLL appears once and $p \geq 1$ levels appear twice due to the aforementioned degeneracy. 
\begin{align}
-\log Z =  f(m) + 2  \sum_{p=1}^{\infty} f( m_p ),   \qquad  m_p \equiv  \Big[m^2 + 2 |eB| p\Big]^{1/2}\,,
\label{LL-sum} 
 \end{align}
where $m_p$ is effective mass associated with level $p$.  The functional form of the contribution of the LLL and higher LLs are the same, and is given by 
\begin{align}
f(m)& =     V_{\mathbb R^3} \frac{|eB|}{4\pi^2}  \int_{-\infty}^{\infty} dk_z  {\rm tr}_{\cal R}   \log  (1+ e^{- \beta\sqrt { k_z^2 + m^2}    } \Omega  )     + {\rm c.c.}\xrightarrow{k_z  = m \sinh t , \;  {\rm Taylor \;expand\; log}}  \cr
&=  V_{\mathbb R^3} \frac{|eB|}{2\pi^2}   m  \int_{0}^{\infty} dt \cosh t   \sum_{n=1}^{\infty} \frac{(-1)^{n+1}}{n}   e^{- n \beta m \cosh t }  \left( {\rm tr}_{\cal R}  \Omega^n 
+ {\rm c.c.}  \right)
\cr
&=  V_{\mathbb R^3} \frac{|eB|}{2\pi^2}   m   \sum_{n=1}^{\infty} \frac{(-1)^{n+1}}{n}    K_1(m \beta n)  \left( {\rm tr}_{\cal R}  \Omega^n 
+ {\rm c.c.}  \right)\,.
\end{align}

It is straightforward to repeat the same steps for fermions endowed with periodic boundary conditions. As a result,  the fermion induced potential for the Wilson line can be expressed as 
\begin{align} 
{\cal V}^{\pm} &=  {\cal V}_{\rm LLL}^{\pm} +  2   \sum_{p=1}^{\infty} {\cal V}_{\rm p^{\rm th}-LL}^{\pm}  \cr 
&=  \left( \frac{|eB|}{2\pi}\right)  \frac{1}{\pi L^2 }    \sum_{n=1}^{\infty} \frac{(\pm)^{n} }{n^2}  (mLn)   K_1(m L n)  \left( {\rm tr}_{\cal R}  \Omega^n + {\rm c.c.}  \right)  \cr
&+ \left( \frac{|eB|}{\pi}\right)  \sum_{p=1}^{\infty}    \frac{1}{\pi L^2 }    \sum_{n=1}^{\infty} \frac{(\pm)^{n} }{n^2}  (m_pLn)   K_1(m_p  L  n)  \left( {\rm tr}_{\cal R}  \Omega^n + {\rm c.c.}  \right) \,.
\label{LL-sum}
\end{align}
The leading term ${\cal V}_{\rm LLL}^{\pm}$   is exactly the  fermion induced term in  (\ref{result for strong field}).  Since the energy of the LLL, $E_{-,p=0, k_z}$, is $B$ independent, the linear behavior with $B$ comes only from the density of states. The terms with $p \geq 1$ are the contributions from the higher Landau levels.  Note that apart from the factor of two difference with respect to the LLL contribution coming from the spectral degeneracy, 
the functional form of these contributions are the same as the LLL with  the replacement $m \rightarrow m_p=  \Big[m^2 + 2 |eB| p\Big]^{1/2}$, where $m_p$ is an effective mass of quarks  associated with level $p$. 

In the large-$B$ limit,  the contributions coming from higher Landau levels are exponentially suppressed, for example,   $K_1(m_pLn)/ 
K_1(mLn)  \sim e^{-L \sqrt{2p |eB|+1  }}$, and we obtain   (\ref{result for strong field}). 

{\bf Equivalence of  (\ref{the main result of the work}) and   (\ref{LL-sum}) :}  To see this, 
we  start with (\ref{the main result of the work})  and use 
 the identity 
 \begin{align} 
 \frac{1}{\tanh x} = 1+ 2 \sum_{p=1}^{\infty} e^{-2 p x} \,.
 \end{align}
 This helps us to 
 express  (\ref{the main result of the work})  as a summation over all Landau levels.  Writing the fermion induced potential as $\sum_{n=1}^{\infty} I^{(n)} (\pm)^n  \left( \tr_{\cal R} \Omega^n + \rm c.c.  \right)$,  we have 
 \begin{align}
I^{(n)} &= \frac{2}{\left(4\pi\right)^2L^4} \int_{0}^{\infty}\frac{dy}{y^3}e^{-\frac{n^2}{4y}-m^2L^2y}\frac{e BL^2y}{\tanh \left(e BL^2 y\right)}  \cr
&=  \frac{ |e B| }{ 8\pi^2L^2} 
   \left( \int_{0}^{\infty} \frac{dy}{y^2}  e^{-\frac{n^2}{4y}-m^2L^2y}   + 2 \sum_{p=1}^{\infty}  \int_{0}^{\infty} \frac{dy}{y^2}   
 e^{-\frac{n^2}{4y}-(m^2 + 2 p |e B|) L^2y}  \right)  \cr
 & =  \left( \frac{|eB|}{2\pi}  \right)  \frac{1}{\pi L^2} \frac{nLm K_1(nLm)}{n^2}   +  \left( \frac{|eB|}{\pi}  \right)   \sum_{p=1}^{\infty}  \frac{1}{ \pi L^2} \frac{nLm_p K_1(nLm_p)}{n^2}   \cr
& = I_{0}^{(n)} +2   \sum_{p=1}^{\infty} I_{p}^{(n)}\,,
\end{align} 
which is  a sum over all Landau levels, equal to  (\ref{LL-sum}).

\subsection{Magnetic susceptibility and its jump across the deconfinement transition}
The magnetic susceptibility is a measure of the response of the QCD thermal equilibrium state (or ground state) to an external magnetic field (see for example   \cite{ Bali:2013esa, Bonati:2013lca}.) 
Here, we  identify the jump in magnetic susceptibility as an order parameter for the confinement/deconfinement phase transition. 
We consider the magnetic susceptibility first for thermal and then for spatial compactification of QCD-like theories.

{\bf Thermal compactification:} Denote the free energy density of QCD as a function of magnetic field and 
inverse temperature  as 
${\cal F} (B, \beta) = -\frac{1}{\beta V_{\mathbb R^3}} \log  Z(B, \beta) $  where $Z$ is the  thermal partition function.
We define the magnetic susceptibility  as: 
\begin{align}
\xi = -  \frac{\partial^2 {\cal F}}{\partial (eB)^2} \Big |_{B=0}\,.
\end{align}
The free energy can be calculated in two related ways. One is by simply extremizing the one-loop potential with respect to holonomy, 
and the other is by using methods of statistical mechanics. Both yield the same result. 

In the  high-temperature deconfined phase, the minimum of the one-loop potential is located at 
$\Omega=1$  and consequently we can use (\ref{the main result of the work}), keeping in mind that ${\cal F}={\cal V}^{\mathbb R^3 \times \mathbb S^1_-}$, to find 
\begin{align}
\xi (\beta< \beta_c)  & = 
\frac{1}{3 \pi^2} \left[\sum_{n=1}^{\infty} K_0(mn \beta) (-1)^{n+1}   \right] {\rm dim}({\cal R})   \cr
& = \frac{1}{3 \pi^2}  \left[ \int_0^{\infty}  dt \frac{1}{e^{m \beta \cosh t } +1}  \right]   {\rm dim}({\cal R}) 
 \approx  
  \left\{ \begin{array}{ll}
 O(N^1)  & \qquad  {\cal R} = F\,,  \cr
 O(N^2)   &  \qquad  {\cal R} =  \mbox{AS/S/Adj}\,, 
\end{array} \right. 
\label{ms-1}
\end{align}
where $\beta_c \sim \Lambda^{-1}$ is the strong length scale.  Clearly, $ \xi >0   $ and the  deconfined phase  is paramagnetic. Also, we find that (\ref{ms-1}) is compatible with the  large-$N$ scaling of  liberated quarks and their free energy.

For the low temperature confined phase, we cannot calculate the magnetic susceptibility due to strong coupling. 
However, there exists  a semi-classically calculable deformation of QCD and YM theory  which is continuously connected to confining low temperature regime \cite{Unsal:2008ch, Shifman:2008ja}. Multiple non-perturbative aspects of deformed QCD confirming the continuity idea are studied in continuum  \cite{  
Ogilvie:2012is, Meisinger:2009ne,  Thomas:2011ee, Zhitnitsky:2012ej, Zhitnitsky:2013hs } and    in lattice   \cite{Vairinhos:2010ha, Vairinhos:2011gv}. 
The main idea is to deform Yang-Mills theory with a center-stabilizing double-trace operator on small $\mathbb S^1  \times \mathbb R^3$ 
such that the minimum of the potential is at a center-symmetric point. 
 For example, for the defining representation, the minimum of the potential is at 
$\Omega=\eta_N \  {\rm Diag} \left( 1, e^{i \frac{2\pi}{N}},  e^{i \frac{4\pi}{N}}, \ldots,  e^{i \frac{2\pi(N-1)}{N}} \right)$,
where $\eta_{\rm odd}=1$  and $\eta_{\rm even} = e^{i \frac{\pi}{N}}$, as shown in Fig.~\ref{Phse diagram}c for $SU(2)$ gauge 
group. (See Section \ref{sec:abel} for the relation between the strong and weak coupling center-symmetric regimes.)

In the weak coupling abelian confinement regime,   the one-loop induced potential for fermions is   still (\ref{the main result of the work}), but the implication is now {\it different}.  The reason is that 
introducing fundamental fermions in the weak coupling confinement regime of deformed Yang-Mills  distorts center-symmetric vacuum only  slightly. In fact, the trace of the Wilson line changes as
$\frac{1}{N} {\rm tr} \Omega =0 \rightarrow \frac{1}{N} {\rm tr} \Omega =O\left( N^{-1} \right)$, i.e. the theory almost respects center symmetry.   
In  the framework of deformed-QCD, which provides  a weak coupling continuation  of the confined phase,  
 we can calculate  the sign and 
$N$ scaling of the magnetic susceptibility. Since the center-symmetry is preserved the quarks are confined in color-singlet states, and therefore we find $ \xi = O(N^0) >0$ and the theory is in a paramagnetic phase. 
It is reasonable to assume that this result in weak coupling abelian confinement regime  extrapolates to strong coupling non-abelian confinement regime. In fact, $O(N^0)$ magnetic susceptibility  is in accordance with the fact that the spectral density of the color singlet states (and free energy density)  in the confined phase is $O(N^0)$.  More explicitly, working with a hadron resonance gas model \cite{Endrodi:2013cs} in the large-$N$ limit, 
we obtain a susceptibility of order $O(N^0)$. 
Therefore, 
 the $N$ scaling differs quantitatively between the   deconfined  and confined phases:  
 \begin{align}
 \xi (\beta) =  \left\{ \begin{array}{ll}
O(N^1) \;  {\rm or} \;   O(N^2)   & \qquad  {\beta < \beta_c} \,, \cr
O(N^0)  &  \qquad  {\beta>\beta_c} \,,
\end{array} \right.
\label{Nc-scale}
\end{align}
 for one-index and two-index representation fermions, respectively. The jump in the  magnetic susceptibility 
provides an  order parameter for deconfinement phase transition.  This jump agrees very well with recent lattice studies \cite{Bonati:2013lca}.

{\bf Spatial compactification:}  We can also study the response of the spatially compactified theory to external magnetic field.  
Define  the ``twisted susceptibility" in the zero temperature, but spatially compactified theory, as 
\begin{align}
{ \xi}^{\rm tw} = -   \frac{ \partial^2}{\partial (eB)^2} \left( - \frac{1}{L V_{\mathbb R^3} }  \log  Z_{+} (B, L) \right)\,.
\end{align}
For ${\cal R} =$ F/AS/S in the small-$L$  regime, $L<L_c$ and $L_c \sim \Lambda^{-1}$,  where spatial (approximate) center symmetry is spontaneousy broken, we have  $\xi^{\rm tw}  \sim O(N^1) $ for F and  $O(N^2) $ for AS/S, and the susceptibility is negative  $\xi^{\rm tw}  < 0$, i.e. the phase $L<L_c$ is diamagnetic.  
On the other hand, for $L>L_c$ we have approximate center symmetry, assuming large $N$ and keeping $n_f$ small, and hence the quarks form singlets and we have $ \xi^{\rm tw}  \sim O(N^{0})<0$.   Note that the signs of susceptibilities are opposite for the thermal versus spatial compactification for complex representations, but the $N$ scaling of  $\xi^{\rm tw}$ is the same as the regular susceptibilities (\ref{Nc-scale}). 
For ${\cal R} =$ Adj with periodic boundary conditions,  there is no center symmetry changing phase transition for sufficiently light fermions, 
and $\xi  \sim O(N^0) $ at any $L$. We explore this case in the next section.

\section{Massive QCD(adj) in external magnetic field}

By inspecting  the fermion induced one-loop potential (\ref{the final form of the effective potential}) and (\ref{summary of all representations}), it is not hard to see that the center symmetry is broken for all representations ${\cal R}$ except for the adjoint representation of $SU(N)$ with periodic boundary conditions (spatial compactification).  
QCD with $n_f$ adjoint Weyl fermions, QCD(adj), possesses  a classical global chiral $SU(n_f)\times U(1)$ symmetry. The $U(1)$ symmetry is anomalous and reduces down to ${\mathbb Z}_{2N n_f}$ due to instanton effects. 
 Below, we restrict attention to $n_f=2$, in which case the global symmetry is just $SU(2) \times {\mathbb Z}_{4N }$.

To couple the system to a $U(1)$  magnetic field, we gauge a $U(1)$ subgroup of the flavor $SU(2)$. This $U(1)$ subgroup is taken to be  of the diagonal form  $\mbox{diag}(1,-1)$. This amounts to assigning opposite charges to  the two  different flavors which in turn guarantees the absence of gauge anomalies. In addition, requesting QCD(adj) to be an asymptotically free theory, we find that $n_f$ has to be either $2$ or $4$. At small compactification radius, $NL\Lambda/2 \pi \lesssim 1$ where $\Lambda$ is the strong coupling scale, 
the  Kaluza-Klein modes as well as the modes which carry a fraction of the KK-momentum  are 
weakly coupled and can be integrated out perturbatively. 
Hence, the one-loop potential resulting from integrating out the non-zero Kaluza-Klein modes of the gauge field and $n_f$ Weyl fermions 
with mass $m$ reads, in the limit of large-magnetic fields, 
\begin{eqnarray}
{\cal V}^{\mathbb R^3 \times S^1_L}_{+} [\Omega]=\frac{2}{\pi^2 L^4}\sum_{n=1}{\cal M}_n^2\left(m,B\right)\frac{\left|\mbox{tr}\Omega^n\right|^2}{n^4}\,, 
\label{spatial-EHGPY bulk} \qquad 
\end{eqnarray}
with effective mass (square)  for the Wilson line 
\begin{eqnarray}
{\cal M}_n^2=-1+\frac{n_f}{4}xz_n^3K_1\left(z_n\right)\,, \quad z_n=nmL\,,  \;\; x=\frac{|eB|}{m^2}.
\end{eqnarray}
The traces $\mbox{tr}\Omega^n$, with $n=\lfloor N/2 \rfloor $  where $n=\lfloor \cdot \rfloor $  is the (lower) floor function, 
 are independent variables.  Therefore, if the effective mass square  ${\cal M}_n^2$ are positive for all $n\leq  \lfloor N/2 \rfloor $,  
 then the $\Z_N$ center symmetry is unbroken with ${\rm tr} \Omega^n =0$ for all  $n \neq 0$ mod $N$. 
 If ${\cal M}_1^2, {\cal M}_2^2, \ldots  $ are negative, then the center symmetry is completely broken.   
 If some of the masses  are tachyonic, then a subgroup of  $\Z_N$ center symmetry breaks down  spontaneously, for details 
 see \cite{Unsal:2010qh}. 

The vanishing of the effective  mass square for $x=\{0,1,5,10\}$ occurs at $z_n^*=\{2.07,2.39,$\newline $5.44,6.61\}$ for $n_f=2$, and at $z_n^*=\{3.16,3.39,6.61,7.67\}$ for $n_f=4$.  Since $z_n^*$ increases with increasing the field strength, a strong field will stabilize the center symmetry for larger values of the compact dimension $L$ at fixed $m$. In effect, this reduces the center-symmetric breaking zone as 
illustrated in Fig. \ref{Phse diagram}.   In the formal infinite magnetic field limit, the  center-breaking phase disappears completely for any fixed value of the fermion mass $m$. 
%
\subsection{Abelian confinement and  large-$N$ volume independence regimes}

In the absence of  magnetic field,   a general $SU(N)$ gauge theory with sufficiently light adjoint fermions, $m<m^*\sim \Lambda$,    (as shown on  the left panel of Fig.~\ref{Phse diagram} for $SU(2)$)
endowed with periodic boundary condition is center-symmetric at any value of the compatification radius $L$: 
$\langle \tr \Omega^n \rangle =0$, $ n \neq 0  \; ({\rm mod } \; N  ) $ 
and exhibits  {\it continuity} in the sense of center-symmetry.  As shown in Fig.\ref{distribution}, 
unbroken center symmetric  holonomy  has different implications depending  on whether the theory is weakly or strongly coupled.  
See next section for more details.  A QCD-like  theory remaining center-symmetric at any compactification radius has two extreme regimes: 
\begin{itemize} 
\item{ $NL\Lambda/2\pi \gg 1$ : non-abelian confinement,  volume independence  (at large $N$) regime.}  
\item{ $NL\Lambda/2 \pi \ll 1 $: abelian confinement, adjoint Higgsing (or Hosotani regime).}
\end{itemize}
The associated Wilson line holonomies are shown in Fig.\ref{distribution} A and C. 

In the absence of magnetic field and for  $m >m^* \sim \Lambda$, there are three regions as shown  in the left panel of Fig.~\ref{Phse diagram} for $SU(2)$.  At sufficiently small-$L$, given by 
$NL m  \lesssim z^*$ for $SU(N)$ (typical values of $z^*$ are given in the previous section), the  $\Z_N$ center symmetry restores   completely  \cite{Unsal:2010qh}.   With mass $m \gtrsim \Lambda$,  the small-$L$ center-symmetric regime  in the lower left corner of the left panel of Fig. \ref{Phse diagram} 
corresponds to  $NL\Lambda \lesssim 1$,  where $L \sim O(N^{-1})$, and is  not the volume independence domain, but rather volume dependent abelian-confinement domain.  On the other hand, for $m \gtrsim \Lambda$ and for $NL\Lambda \gtrsim 1$ (at large $N$) we have a large-$L$ center-symmetric, non-abelian confinement and volume independent regime. These two regimes ($NL\Lambda \lesssim 1$ and $NL\Lambda \gtrsim 1$) are separated  by an intermediate phase in which center-symmetry is spontaneously broken as shown  in Fig.~\ref{Phse diagram} for $SU(2)$.

This phase separation between the large and small $L$ regimes can be avoided in the presence of a strong  magnetic field since the  field sets a new scale which parametrically  enhances the effect of the  adjoint fermions.   In the presence of a very large magnetic field, the condition for  the preservation of 
center-symmetry   is $NLm< z^*\left(\frac{|eB|}{m^2}\right)$ (typical values of $z^*$ in the presence of strong field are given in the previous section) .   
Thus, the hierarchy 
\begin{align}
\sqrt {eB} \gg m > \Lambda
\label{scaling1}
\end{align}
can help the stabilization of the center symmetry  at larger values of $L$, reducing the region in which the center is broken. 
 In particular, in the (formal) exponentially large-$B$ field limit, (such that it can undo the effect of the mass term for fermions)    the intermediate regime in which center-symmetry 
is broken shrinks, and gradually disappears.
Consequently, the infinite-$B$ theory  with any finite  fermion mass $m$ and  large $N$  possesses both volume independent 
non-abelian confinement regime  $NL\Lambda/2\pi \gg 1$ and volume dependent 
abelian confinement regime  $NL\Lambda/2\pi \lesssim 1$, as we vary $L$. In particular, the center symmetry is always respected and it has a double life: one at small $L$ (weak coupling), and the other at large $L$ (strong coupling).

 For the purpose of the center-symmetry preservation in the weak coupling regime, this is a {\it non-decoupling} of large mass fermion, (due to its enhancement by LLL density of states), and  the center-symmetry stabilizes. However,  in the same time the  low energy effective field theory (the dynamics at distances larger than $1/m$) is a pure YM theory up to $\Lambda/m$ corrections!

%
\section{Comments on lattice  realization of abelian confinement  and Hosotani mechanism on $\R^3 \times S^1$}
%
\label{sec:abel}

As already mentioned, it is crucial  to emphasize  that unbroken center symmetry with  
\begin{align}
\langle \tr \Omega^n \rangle =0, \qquad  n \neq 0  \; ({\rm mod }N  )   \qquad {\rm at \; any} L
\label{vanish}
\end{align}
 has {\it multiple   different realizations} depending 
on whether the theory is weakly or strongly coupled.  This difference is not sufficiently addressed in literature, 
the first discussion of it is in  \cite{Unsal:2008ch} and a more through discussion can be found in Section 5 of  
\cite{Poppitz:2012nz}.  Our goal is not to repeat the same argument here, but rather to point out the lattice realization of  the regimes
shown in Fig.\ref{distribution}: 
\begin{itemize}
\item{ $NL\Lambda/2\pi \gg 1$:  Strong coupling non-trivial holonomy, gauge symmetry unbroken, Fig.\ref{distribution}A }
\item{ $NL\Lambda/2 \pi \ll 1 $:  Weak coupling non-trivial holonomy, gauge symmetry broken, Fig.\ref{distribution}C}
\end{itemize}

   \begin{FIGURE}[ht]
    {
    \parbox[c]{\textwidth}
        {
        \begin{center}
        \includegraphics[angle=0, scale=.35]{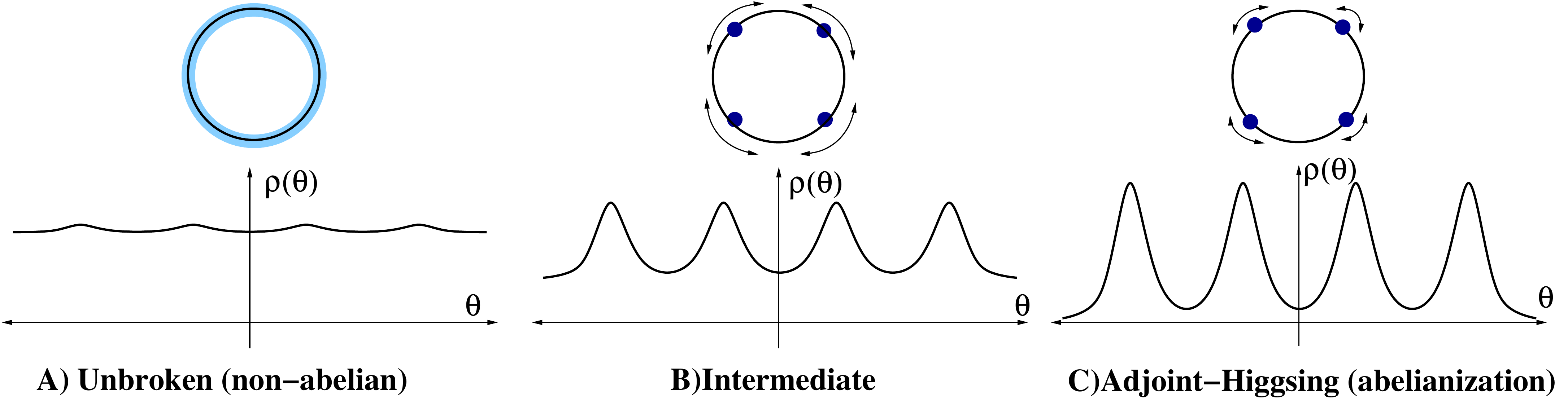}
        \caption 
      { {\it Realizations} of unbroken center symmetry $ \langle \tr \Omega\rangle  =0$ from strong to weak coupling, both in continuum and in lattice.   A) Strong coupling non-trivial holonomy, eigenvalues are randomized over the eigenvalue circle. B) Intermediate coupling. 
      C) Weak coupling non-trivial holonomy. Eigenvalues are at the roots of unity (up to a phase) and their fluctuations are small. 
   These regimes are continuously connected in the sense of center symmetry, C) is non-perturbatively calculable.          
			}
      \label{distribution}
       \end{center}
        }
    }
\end{FIGURE}

In the strong coupling regime, eigenvalues are 
randomized over the dual  circle.  This configuration  cannot be viewed as a minimum of a potential in a local effective field theory, i.e, there is no parametric separation of scales that justify an effective field theory.   In this regime, the average Wilson line determines the free energy.   This is opposite to what happens in the weak coupling (abelian)  confinement regime where Wilson line potential can be viewed as a potential in a local effective field theory (with appropriate parametric separation of scales).

In  center-symmetric weak coupling regime,  (\ref{vanish})   implies that 
the minimum of the one-loop potential for the Wilson line  is at  
\begin{align}
\Omega= \eta_{\rm N}   \left( \begin{array} {ccccc} 
   1 &&&& \\ 
   &e^{i \frac{2\pi}{N}} && & \\ 
 &&     e^{i \frac{4\pi}{N}} & &  \\
 &&& \ddots & \\
&&&&  e^{i \frac{2\pi(N-1)}{N}} 
  \end{array} 
   \right)\,, \qquad {\rm where}  \;\;   \left\{ \begin{array}{l}
 \eta_{\rm odd}=   1  \qquad \cr 
   \eta_{\rm even}=  e^{i \frac{\pi}{N}}\,, 
   \end{array} \right.
\label{min-pot} 
\end{align}
 as shown in Fig.~\ref{Phse diagram}c for $SU(2)$   and Fig~\ref{distribution}C for $SU(4)$
 gauge 
groups.  In this regime, because of the weak coupling, the fluctuation of the eigenvalues are small and the theory, to all orders in perturbation theory, undergoes adjoint Higgsing, i.e. the long distance theory abelianizes:
\begin{align}
SU(N) \rightarrow U(1)^{N-1}.  
\end{align}
  The abelianized regime is realized  if the theory is weakly coupled at the scale of  the   inverse  of the lightest W-boson mass, $m_W^{-1} = LN/2\pi $ for the center-symmetric 
   background \cite{Unsal:2008ch},\footnote{If $N$= few, the appearance of $N$ in   (\ref{abelianization}) hardly matters. However, in the large-$N$ limit, the correct combination determining if a center-symmetric theory is weakly coupled or not is  $ \frac{LN\Lambda}{2 \pi} $.}
   \begin{align}
   \frac{ g^2 N (LN) }{4\pi} \ll 1  \qquad  {\rm or}  \qquad
   \frac{LN\Lambda}{2 \pi} \lesssim 1 \,. 
   \label{abelianization}
   \end{align}

Whether the   gauge fluctuations (photons of $U(1)^{N-1}$)  which are  massless to  {\it all orders} in perturbation theory  acquire a 
  dynamical mass or not depends on the details of the theory.  
 In deformed-YM,  the photons acquire a mass  via monopole-instanton mechanism \cite{Unsal:2008ch}, and   in  ${\cal N}=1$ SYM and   QCD(adj),   they do so  via the  magnetic bion  mechanisms \cite{Unsal:2007jx}.  
 However, in  ${\cal N}=2$ SYM  in center-symmetric background (\ref{min-pot}) in its Coulomb branch associated with Wilson line,    the photons do not acquire a  dynamical mass.    Despite the fact that monopole-instantons {\it do exist}, their fermion zero mode structure and   ${\cal N}=2$  extended supersymmetry does not permit the generation of mass gap \cite{Seiberg:1996nz}. 
   The weak coupling regime  provides an example of gauge symmetry breaking (or Hosotani mechanism) to  all orders in perturbation theory  for deformed YM and QCD(adj),    and  a  non-perturbative realization of gauge symmetry breaking in the  ${\cal N}=2$ SYM.  
    In the  deformed YM and QCD(adj), the  IR-theory acquires  a mass gap for gauge fluctuations, while in  the  ${\cal N}=2$ SYM,   the  IR theory is gapless   $U(1)^{N-1}$ theory  non-perturbatively.

   The realization of the abelianization regime   in lattice gauge theory requires the mapping of the regime (\ref{abelianization}) to lattice units.   To emulate $\R^3 \times  \S^1$, consider a 4d lattice  $\Lambda_4$ with size $L_1= \Gamma_1 {\mathfrak a}  =L_2  = \Gamma_2 
   {\mathfrak a}  = L_3= \Gamma_3  {\mathfrak a} \gg L_4= \Gamma_4 {\mathfrak a} $   
   where $\Gamma_\mu$ is the number of sites in a given direction, and ${\mathfrak a}$   is lattice spacing. 
      This is an asymmetric discretized 4-torus.
    Define  lattice gauge action with adjoint fermions  as 
    \begin{align}
    S[U]= \beta \sum_{p \in \Lambda_4}  \frac{1}{N} (\tr U[ \partial p]  +  \tr U^{\dagger}[ \partial p] ) + S_{\rm fermion}\,,  \qquad {\rm where} \; \; 
    \beta= \frac{ g_0^2 N }{4\pi}, 
    \end{align}
 and   $  g_0^2 = g^2({\mathfrak a})$
   is the bare coupling constant at the lattice cut-off scale ${\mathfrak a}$. In order to achieve abelianization of the long distance dynamics,  one needs weak coupling at the scale  $L_4N = N \Gamma_4 {\mathfrak a}$,  
      \begin{align}
\frac{ g^2 N ( N \Gamma_4 {\mathfrak a} )}{4\pi} \ll 1   \qquad L_4 \equiv \Gamma_4 {\mathfrak a} = {\rm fixed}, \;\; {\rm as}  \;\;   \Gamma_4 \rightarrow \infty,  \;\; {\mathfrak a} \rightarrow 0\,.
   \label{lattice-abelianization}
   \end{align}
 Once this is achieved, the dynamics  abelianizes at distances larger than the  inverse lightest W-boson mass, where 
 \begin{align} 
 m_W = \frac{2 \pi}{ L N} \qquad {\rm  in \;  continuum},  \qquad \qquad    m_W =  \frac{2}{ {\mathfrak a}} \sin \frac{  \pi}{ \Gamma_4 N} \qquad {\rm  in \;  lattice} \,.
 \end{align}   
 For two point connected correlators, $\langle O({\bf x}) O({\bf 0})  \rangle$, in order to disentangle the short distance degrees of freedom from the long-distance $U(1)^{N-1}$  photon modes, one needs  (along the non-compact directions)   separations 
 larger than  $|{\bf x}| \gtrsim  \frac{ \Gamma_4 N   {\mathfrak a}        }{2 \pi}$. 
 Therefore, to see the  abelianized dynamics of the gapless photons (in perturbation theory), one must have 
 \begin{align}
  L_i = \Gamma_i  {\mathfrak a}   \gtrsim   \frac{ \Gamma_4 N  {\mathfrak a}   }{2 \pi} =   \frac{ L_4 N  }{2 \pi}  \qquad  {\rm abelianized \;\;  (Hosotani) \;\;  regime}\,.
  \label{H-R}
      \end{align} 
This may be considered as the {\it Hosotani regime}  of the lattice gauge theory  formulated on $T^3 \times S^1$. It is extremely  important to note that  $L_i  \gtrsim   L_4 $ is {\it not sufficient} to see the Hosotani regime.  The decoupling of the non-Cartan sub algebra degrees of freedom i.e., W-bosons, occurs at scales larger than  $m_W^{-1} \sim   L_4 N  $.  In particular,  at large-$N$ limit, the abelianization only occurs  at $L_i=\infty $ regardless of how small $L_4$  is so long as it is $O(N^0)$.

\subsection{The resolution of Eguchi-Kawai versus Hosotani puzzle} 
Both Hosotani mechanism and Eguchi-Kawai demands  the very same {\it unbroken center symmetry} condition in QCD(adj), yet they are completely different physical phenomenon.  This is what we mean by  Eguchi-Kawai versus Hosotani puzzle.  

The overall picture and resolution should  now  be clear. 
In a working  Eguchi-Kawai reduction,  center symmetry does  not break, and consequently, in the large-$N$ limit, gauge symmetry never breaks regardless of how small $L_4$ is so long as it is  $O(N^0)$.  
 In  Hosotani mechanism, center symmetry does  not break either, and 
yet gauge symmetry breaks at sufficiently weak coupling, which, in the large $N$ limit, scales as   $L_4 \sim O(N^{-1})$.
This   is how {\it i)} abelianized (Hosotani) regime  where gauge symmetry is broken, {\it ii)}  non-abelian volume independence (Eguchi-Kawai) regime  and   {\it iii)}  non-abelian large-$L$, finite-$N$ regimes  where gauge symmetry remains unbroken
   mutually exclude each other without leading to any contradiction.

This intricate working of the physical scales is also most likely the reason that the  research along these two directions  (despite  relying on the same physical condition of unbroken center symmetry) remained mutually exclusive so far.  Clearly, without careful deliberation of scales  they are in apparent conflict  with each other.

\subsection{How large  should the box be in order to see the setting of  mass gap and abelian confinement?}
Non-perturbatively, we also know that the photons on $\R^3 \times \S^1 $ acquire a mass gap in deformed YM   
and QCD(adj) with heavy fermions  via monopole-instanton  mechanism and in QCD(adj) with massless or light fermions via the  magnetic bion mechanism \cite{Unsal:2008ch, Unsal:2007jx} . 
This gap is, for example, in weak coupling deformed YM or center-symmetric regime of massive QCD(adj) is given by $m_{\rm gap}  = m_W e^{ - \frac{ 4 \pi^2 }{g^2(m_W) N} }$  =   $\Lambda  (\Lambda L N)^{5/6} $  where $\Lambda$ is the strong scale of YM theory.

In order to see the gap for the (dual) photons,  the box size must also be larger than the inverse of the mass gap; otherwise  one will always 
{\it erroneously} conclude that the theory is gapless.\footnote{This is the main  danger with  lattice  simulations of the abelian confinement regime. Although one can see  (by current techniques) both abelianization  (Hosotani regime) and gapless photons, since no dramatic  hierarchies are required to achieve this but just (\ref{H-R}),  it is probably fairly hard to demonstrate  the appearance of the mass gap for gauge fluctuations.}  
 This requires 
\begin{align}
&L_i \gtrsim m_{\rm gap}^{-1} = m_W^{-1}   e^{ +\frac{ 4 \pi^2 }{g^2(m_W) N} }   =   \frac{ L_4 N  }{2 \pi} e^{ +\frac{ 4 \pi^2 }{g^2(m_W) N} }\,.   \qquad 
 {\rm abelian\;\;  confinement \;\;  regime}
\end{align}
Admittedly, it may be difficult  to achieve such a hierarchy in practical simulations and  also hard   to see the regime of abelian confinement, but   we are not pessimists on this, and  there is a very strong  incentive to pursue this direction, see Section~\ref{incentive}.

On a practical side, on a $ \Gamma_i^3 \times \Gamma_4 = 16^3 \times 4$ lattice formulation of  $SU(3)$ lattice QCD(adj) with $m {\mathfrak a}  =0.1$ where $m$ is bare quark mass,  the small-$L$ confined phase is achieved at  $\beta > \beta^* = 6.30$ \cite{Cossu:2009sq}.  Making, for example,  $\beta \gtrsim 10$, forcing the theory to remain 
weakly coupled at  $N \Gamma_4{\mathfrak a} $, one can certainly achieve abelianization.  But making $\beta$ so large also makes the  
length scale of  the mass gap $m_{\rm gap}^{-1}$ much larger than  box-size $\Gamma_i {\mathfrak a} $ along the large dimensions. This, as explained above,  will lead to the  incorrect conclusion that     the theory is gapless.   Thus, one  needs to make $ \Gamma_i $  
as large as possible and  make $\beta>\beta^*$ as small as possible while remaining in  abelianized regime.

\subsection{Why  is the  weak coupling corner     important both for  lattice and continuum studies?} 
\label{incentive}
There are only a handful of theories in which confinement and mass gap  can be understood by  reliable field theory methods in three and four dimensions. These are:
\begin{itemize}
\item{ Softy broken ${\cal N}=2$ SYM theory down to ${\cal N}$=1 SYM on $\R^4$ and $ \R^3 \times \S^1$ \cite{ Seiberg:1996nz}}, 
{\item Polyakov model on  $\R^3$ \cite{Polyakov:1976fu} }
{\item QCD(adj)  \cite{Unsal:2007jx} and deformed YM  \cite{Unsal:2008ch} on small $ \R^3 \times \S^1 $ }. 
 \end{itemize} 
 
  It is currently not feasible  to simulate softy broken ${\cal N}=2$ SYM  theory on lattice despite much progress in  lattice supersymmetry.
   It is also technically very difficult to simulate Polyakov model on $\R^3$, due to fine tunings (for scalar masses and quartics,  for example)  required to reach the continuum limit. Neither of these difficulties   are present in  QCD(adj) with massive fermions  and deformed YM, while   the  problem of accessibility of the abelian confinement regime is present in all three cases.    
 It seems to us that  QCD(adj) with massive fermions and deformed YM on small $ \R^3 \times \S^1 $  are a target of opportunity, both of which can easily be simulated.   If abelian confinement regime can be reached,  it   may become an important playground  for both lattice and analytical studies of non-perturbative physics.    The abelian confinement regime  has a potential to help both fields alike. If it can be achieved, it will be the first confrontation of reliable analytical methods against the reliable lattice methods.

\section{Conclusion and future work}
Our main results are: 
\begin{itemize}
\item{At sufficiently large magnetic fields, the fermion induced one-loop potential for Wilson line holonomy undergoes dimensional reduction by two-dimensions. The fermion contribution is enhanced by the density of state of the lowest Landau level.}
\item{ For massive adjoint fermions endowed with periodic boundary condition, changing magnetic field can alter the phase of the theory from a center-broken phase to a center-symmetric phase. This is an exotic phase transition induced by the competition between center-destabilizing  one-loop gauge contributions  and center stabilizing LLL-adjoint fermion contribution.}
\item{ The fully center stabilized theory has both abelian confinement regime and non-abelian confinement regime. These two regimes are continuously connected in the sense of center symmetry, but the behavior of Wilson line eigenvalues is drastically different as shown in 
Fig.~\ref{distribution}}. 
\item{Realizing the abelianization (adjoint Higgsing) in lattice simulations  requires the lattice version of the scaling  
$\frac{L_4N\Lambda}{2\pi}   \lesssim 1$ and  $ L_i \gtrsim  \frac{L_4N}{2\pi}  $,  and is currently feasible \cite{Cossu:2009sq, Cossu}. But 
realizing the setting of abelian confinement regime requires an exponential  hierarchy of scales,  $L_i \gtrsim  \frac{ L_4 N  }{2 \pi} e^{ +\frac{ 4 \pi^2 }{g^2(m_W) N} }     $   on a (physical) size  $L_i^3 \times L_4$ 4-torus (emulating $\R^3 \times \S^1$).   This may be technically challenging, but  is a worthy endeavor because  of questions such as confinement and mass gap in 4d non-abelian gauge theories. This is the first confrontation of reliable semi-classical methods  against  numerical lattice  simulations. }
\end{itemize}

For future work, we aim to study the deformed QCD with light fermions  in the presence of large-magnetic fields,
in the scaling regime $\sqrt {eB} \gg \Lambda \gg m$. 
 \begin{itemize}
\item{ 
  It is already known that confinement and discrete chiral symmetry breaking  can take place at weak coupling as well. 
Our goal is to construct a calculable theories  in which both confinement and non-abelian continuous chiral symmetry breaking take place at weak coupling, and the dynamics is continuously connected to the one on $\R^4$.  This may provide a useful laboratory for QCD on $\R^4$.}
\item{  We would like to understand the role of the magnetic field on the fermonic zero and quasi-modes of monopole-instantons, and bions. These defects are exponentially more important than the 4d instantons.   We would like to understand how large-magnetic fields may alter  
index theorems for monopole-instantons, confinement mechanism, and fermion induced pairing mechanism of (chromo)-magnetic bions. }
\item{ We aim to study the effect of monopole-instantons and the sphalerons associated with monopole-instantons on the chiral magnetic effect.  
It is natural to expect that if 4d instantons induce a chiral magnetic current or non-vanishing fluctuations, then  monopole-instantons effects should enhance that by an exponential amount. }
\end{itemize}

\acknowledgments

It is a pleasure to thank  G. Cossu, G. Dunne,  G. Basar, F. Bruckmann, D. Kharzeev,  and E. Poppitz  for helpful comments.
 The work of M.\"U. is supported in part by DOE grant DE-FG02-12ER41806. The work of M.A. is supported by NSERC Discovery Grant of
Canada.

\vskip+1pc

\end{document}